\documentstyle[aps,floats,epsfig]{revtex}

\begin{document}              

\input epsf
\renewcommand{\topfraction}{1.0}
\twocolumn[\hsize\textwidth\columnwidth\hsize\csname 
@twocolumnfalse\endcsname

\title{A depression before a bump in the highest 
energy cosmic ray spectrum}
\author{L. A. Anchordoqui, 
M. T. Dova and L. N. Epele}
\address{ Departamento de F\'\i sica, Universidad Nacional de La Plata\\
 C.C. 67, (1900) La Plata\\
\it Argentina\\ }
\author{J. D. Swain}  
\address{ Department of Physics, Northeastern University\\
 Boston, Massachusetts, 02115\\ 
 USA}
\maketitle
\begin{abstract}
We re-examine the interaction of ultra high energy nuclei with the 
microwave background radiation. We find that the giant dipole resonance 
leaves a new signature in the differential energy spectrum of iron sources 
located around 3 Mpc: A depression before the bump which is 
followed by the expected cutoff. 

\noindent {\it PACS number(s):} 96.40, 13.85.T, 98.70.S, 98.70.V
\end{abstract}

\vskip2pc]

%%%%%%%%%%%%%%%%%%%%%%%%%%%%%%%%%%%%%%%%%%%%%%%%%%%%%%%%%%%%%%%%%%%%%%%%%%%%%
\section{Introduction}
%%%%%%%%%%%%%%%%%%%%%%%%%%%%%%%%%%%%%%%%%%%%%%%%%%%%%%%%%%%%%%%%%%%%%%%%%%%%%

In 1966 Greisen, Zatsepin and 
Kuz'min \cite{G,Z} noted that the microwave background radiation
(MBR) makes the universe opaque 
to cosmic rays of sufficiently high energy, yielding a steep drop
in the energy cosmic ray spectrum at approximately $ 5
\times 10^{19}$ eV (GZK cutoff). More recently, a fresh interest in
the topic has been rekindled since several extensive air 
showers have been observed which
imply the arrival of cosmic rays with energies above $10^{20}$ eV. 
In particular, the Akeno Giant Air Shower Array (AGASA) 
experiment recorded an event
with energy 1.7 - 2.6 $\times 10^{20}$ eV \cite{Yoshi,Hasha},  
the Fly's Eye experiment reported the highest energy cosmic ray  
event ever
detected on Earth, with an energy 2.3 - 4.1 $\times 10^{20}$ eV
\cite{Bird1,Bird2}, both events being well above the GZK cutoff. 
Deepening the mystery, the identification of the primary 
particle in these showers is still uncertain. On the one hand, the Fly's Eye 
group claims that there is evidence of a transition from a spectrum 
dominated by heavy nuclei 
to one of a predominantly light composition \cite{Bird1}, while 
on the other hand, 
it has also been suggested that a medium mass nucleus also fits the 
shower profile of the highest energy Fly's Eye event \cite{H}. 
In addition, there is 
an unexpected energy gap 
before these events. Although heavy nuclei can be accelerated to high
terminal energies  by  ``bottom up'' mechanisms,
one should note that, for energies above 100 EeV the range of 
the corresponding sources
is limited to a few Mpc \cite{JCronin}. Sigl and co-workers \cite{Sigl} have 
analysed the structure of the 
high energy end
of the cosmic ray spectrum. They found that most 
``bottom up''
models can be ruled out except for
those involving a nearby source, which is 
consistent
with data at the  1$\sigma$ level. Their argument for this is that a 
nearby source can account for the ultrahigh energy events but would also 
produce events in the apparent gap in data obtained to date. In this 
direction, 
Elbert and Sommers have suggested that the highest energy event recorded by
Fly's Eye, could have been accelerated in the neighborhood of M82, which is 
around 
3 Mpc away \cite{ES,W}. 
In relation to the aforementioned possibilities,
we have re-examined the interaction of ultrahigh energy nuclei with the 
microwave background radiation and we have found a new feature
in the ultrahigh energy cosmic ray spectrum  from iron sources located 
around 3 Mpc 
which forms the motivation for the present article. 

%%%%%%%%%%%%%%%%%%%%%%%%%%%%%%%%%%%%%%%%%%%%%%%%%%%%%%%%%%%%%%%%%%%%%%%%%%%%%
\section{Energy attenuation length of ultrahigh energy nuclei}
%%%%%%%%%%%%%%%%%%%%%%%%%%%%%%%%%%%%%%%%%%%%%%%%%%%%%%%%%%%%%%%%%%%%%%%%%%%%%

The energy losses that extremely high energy nuclei suffer during 
their trip to the Earth are due to their interaction with the
low energy photons of the MBR which they
see as highly blue-shifted. The interaction with other radiation
backgrounds (optical and infrared) can be safely neglected for
nuclei with Lorentz factors above $2 \times 10^9$. Although the 
interactions of extremely high energy nuclei with
the relic photons lead to step-by-step energy loss (which needs to be
included in a transport equation as a collision integral), in what
follows we use the continuous energy loss
approximation assuming straight line propagation 
which is reasonable for the energies and distances under consideration
in this paper. 
The relevant mechanisms for energy losses are
photodisintegration and hadron photoproduction (which has a
threshold energy of $\approx 145$ MeV,
equivalent to a Lorentz factor of $10^{11}$, above
the range treated in this article) \cite{BLUE}. 

The disintegration rate of a nucleus of mass $A$ with the subsequent 
production of $i$ nucleons is given by the expression \cite{Ste69},
\begin{equation}
R_{Ai} = \frac{1}{2 \Gamma^2} \int_0^{\infty} dw \,
\frac{n(w)}{w^2} \, \int_0^{2\Gamma w} dw_r
 \, w_r \sigma_{Ai}(w_r)
\label{rate}
\end{equation}
where $n(w)$ is the density of photons with energy $w$ in the 
system of reference in which the microwave background is at 3K and
$w_r$ is the energy of the photons in the rest frame of the nucleus.
As usual, $\Gamma$ is the Lorentz factor and $\sigma_{Ai}$
is the cross section for the interaction. Using the expressions for the cross 
section fitted by Puget {\it et al.} \cite{PSB}, it is possible to work out an
analytical solution for the nuclear disintegration rates \cite{sudaf}. 
After summating them over all the possible channels for a given 
number of nucleons one obtains the effective nucleon loss rate. 
The effective $^{56}$Fe 
nucleon loss rate obtained after carrying out 
these straightforward but rather lengthy steps
can be parametrized by,
\begin{mathletters}
\begin{equation}
R(\Gamma)=3.25 \times 10^{-6}\, 
\Gamma^{-0.643}                                          
\exp (-2.15 \times 10^{10}/\Gamma)\,\, {\rm s}^{-1} 
\end{equation}
if $\Gamma \,\in \, [1. \times 10^{9}, 3.68 \times 10^{10}]$, and 
\begin{equation}
R(\Gamma) =1.59 \times 10^{-12} \, 
\Gamma^{-0.0698}\,\, {\rm s}^{-1}   
\end{equation}
if $ \Gamma\, 
\in\, 
[3.68 \times 10^{10}, 1. \times 10^{11}]$. 
\end{mathletters}
It is noteworthy that knowledge of the iron 
effective nucleons loss rate alone is enough to obtain the corresponding 
value of 
$R$ for any other nuclei \cite{PSB}.

The emission of nucleons is isotropic in the
rest frame of the nucleus, and so the averaged fractional
energy loss results equal the
fractional loss in mass number of the nucleus, viz., the Lorentz
factor is conserved. 
The relation which determines
the attenuation length for energy is then, assuming an initial iron nucleus, 
\begin{equation}
E = E_g \,\, e^{-R(\Gamma ) \, t / 56}
\label{constraint}
\end{equation}
where $E_g$ denotes the energy with which the nuclei were 
emitted from the source, and $\Gamma = E_g / 56$.

\begin{figure}
%\label{} 
\centering 
\leavevmode\epsfysize=8cm \epsfbox{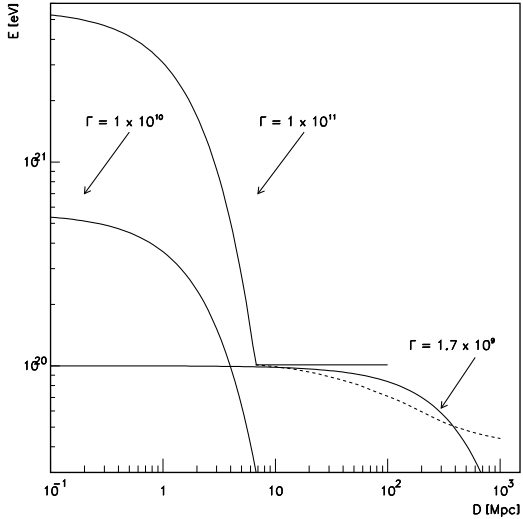}\\ 
\caption{Energy of the surviving nuclei vs. propagation
distance. It is also included the energy attenuation length of the
surviving nucleon (dot line).}
\end{figure} 

In Fig. 1 we have plotted the total energy of the heaviest surviving
fragment as a function of the distance for initial iron nuclei. 
Note that the values obtained here are consistent with the
ones obtained by Cronin using Monte Carlo simulation \cite{JCronin}.
One
can see that nuclei with Lorentz factors above $10^{10}$ cannot survive 
for more
than 10 Mpc \cite{Ste97}.

\begin{figure}
%\label{} 
\centering 
\leavevmode\epsfysize=8cm \epsfbox{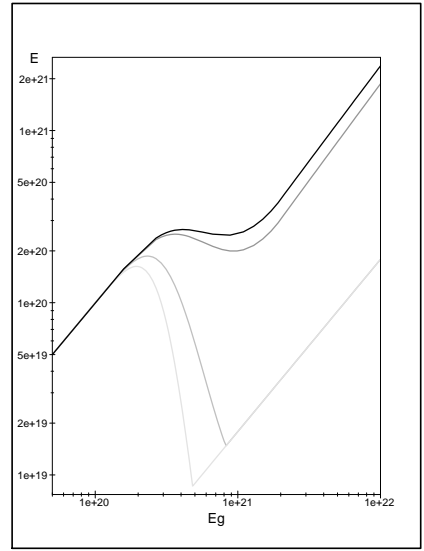}\\ 
\caption{Relation between the injection energy of an iron nucleus 
and the 
final energy of the 
photodisintegrated nucleus for different values of the propagation distance
(from grey to black 20 Mpc, 10 Mpc, 3.5 Mpc, 3 Mpc).}
\end{figure} 

In Fig. 2 the relation between the injection
energy and the energy at a time $t$ for different propagation distances 
is shown. The graph indicates that the final energy of the nucleus is
not a monotonic function. It has a maximum at a critical energy
and then decreases to a minimum before rising again as $\Gamma$
rises, as was first pointed out by Puget {\it et al.} \cite{PSB}. 
The fact that the energy $E$ is a multivaluated function of $E_g$
leads to a pile-up in the
energy spectrum. Moreover, this
behaviour enhances a hidden feature of the energy
spectrum for sources located beyond 2.6 Mpc: A depression that preceeds 
a bump
that would make the events at the end of the spectrum (just before the cutoff)
around 50\% more probable than those in the depressed region. 
To illustrate this, 
let us discuss the 
evolution of the differential energy spectrum of nuclei.

%%%%%%%%%%%%%%%%%%%%%%%%%%%%%%%%%%%%%%%%%%%%%%%%%%%%%%%%%%%%%%%%%%%%%%%%%%%%
\section{Modification of the cosmic ray spectrum}
%%%%%%%%%%%%%%%%%%%%%%%%%%%%%%%%%%%%%%%%%%%%%%%%%%%%%%%%%%%%%%%%%%%%%%%%%%%%

The photodisintegration process
results in the production of nucleons of ultrahigh energies
with the same Lorentz factor of the parent nucleus. 
As a consequence, the total number of particles is
not conserved during propagation. However, the solution of the problem
becomes quite simple if we separately
treat both the evolution of the heaviest fragment and those fragments corresponding 
to nucleons emitted from the travelling nuclei.
The evolution of the differential spectrum of the 
surviving fragments is governed by a
balance equation 
that takes into account the conservation of the
total number of particles in the spectrum.
Using the formalism presented by the authors in reference \cite{nos}, 
and considering the case of a single  source located at $t_0$ 
from the observer, with 
injection spectrum $Q(E_g, t) \,= \,\kappa \, E_g^{- \gamma} \,
\delta(t - t_0)$,  
the number of particles with energy $E$ at time $t$ is given by,
\begin{equation}
N(E, t) dE  = \frac{\kappa E_g^{-\gamma+1}}{E} dE, 
\label{espectro}
\end{equation}
with $E_g$ fixed by the constraint (\ref{constraint}).

Let us now consider the evolution of nucleons generated by 
decays of nuclei during their propagation. 
For Lorentz factors less than  $10^{11}$
and distances less than 100 Mpc the energy with which the secondary
nucleons are produced is approximately equal to the energy with which they
are detected here on Earth. The number of nucleons with energy $E$ at time
$t$ can be approximated by the product of the number of nucleons generated
per nucleus and the number of nuclei emitted.
When the nucleons are emitted with energies above 100 EeV
the losses by meson photoproduction start to become significant.
However, these nucleons come from heavy nuclei with 
Lorentz factors $\Gamma  > 10^{11}$ which are completely disintegrated
in distances of less than 10 Mpc. Given that the mean free path of the 
nucleons is about $\lambda_n \approx 10$ Mpc, it is reasonable to 
define a characteristic time   $\tau_{_{\Gamma}}$ given by the
moment in which the number of nucleons is reduced to $1/e$ of its 
initial value $A_0$. In order to determine the modifications of the spectrum 
due to the losses which the nucleons suffer due to interactions with the relic
photons, we assume that the iron nucleus emitted at $t = t_0$ 
is a travelling source which at the end of a time  $\tau_{_{\Gamma}}$
has emitted the 56 nucleons together. In this way the injection 
spectrum of nucleons ($\Gamma \approx 10^{11}$) can be approximated by,
\begin{equation}
q(E_G,t) =  
\kappa \, A_0^{-\gamma+1} \, E_G^{-\gamma} \delta(t - \tau_{_{\Gamma}}),  
\end{equation}
where $A_0$ is the mass of the initial nucleus and the energy with
which the nucleons is generated is given by
$E_G = E_g / A_0$.

\begin{figure}
%\label{} 
\centering 
\leavevmode\epsfysize=8cm \epsfbox{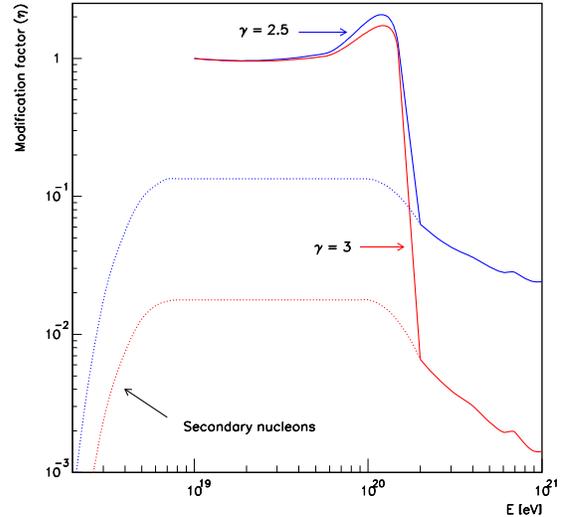}\\ 
\caption{Modification factors for sources of iron nuclei at 20 Mpc together
with the spectra of secondary nucleons.}
\end{figure}

The number of nucleons with energy $E$ at time $t$ is 
given by,
\begin{equation}
n(E,t) dE = \frac{\kappa \, A_0^{-\gamma+2} \, E_G^{-\gamma+1}}{E} dE   
\end{equation}
and the relation between injection energy and the energy at time $t$ remains
fixed by the relation, $\, A \, (t - \tau_{_{\Gamma}}) \, - \,  
{\rm{Ei}}\,(B/E) 
+ \, {\rm{Ei}}\, (B/E_G) 
= 0
$,  Ei being the exponential integral,
and $A$, $B$ the parameters of the fractional energy loss of 
nucleons previously fitted by the authors \cite{nos}. 

The modification factor $\eta$, is defined as the ratio between the
modified spectrum and the unmodified one. In Fig. 3 we plot the
modification factors for the case of sources of iron nuclei (propagation
distance 20 Mpc) together with the spectra of secondary nucleons. It is
clear that the spectrum of secondary nucleons around the pile-up is
at least one order of magnitude less than the one of the surviving
fragments.
In Figures 4 and 5 we have 
plotted the modification factor
for different propagation distances around 3 Mpc. 
They display a bump and a cutoff and, in addition, 
a depression before the bump. It is important to stress that the
mechanism that produces the pile-up which can be seen
in Figures 3, 4 and 5 is completely different to the one that produces
the bump in the case of nucleons.

In this last case, the photomeson production involves 
the creation of new particles that carry off energy yielding
nucleons with energies ever closer to the photomeson production threshold. 
This mechanism, modulated by the fractional energy loss, is responsible for
the bump in the spectrum.
The cutoff is a consequence of the conservation of the number
of particles together with the properties of the injection spectrum 
($\int_{E_{_{\rm th}}}^\infty 
E_g^{-\gamma} dE_g< \infty$). 

\begin{figure}
%\label{} 
\centering 
\leavevmode\epsfysize=4.8cm \epsfbox{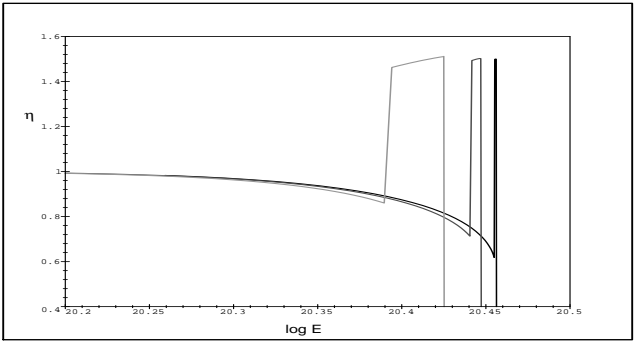}\\ 
\caption{Modification factor of single-source energy 
espectra for different values of propagation distance (from grey to
black 3 Mpc, 2.7 Mpc and 2.6 Mpc) assuming a
differential  
power law injection spectrum with spectral index $\gamma = 2$.}
\end{figure} 

In the case of nuclei, since the
Lorentz factor is conserved, the surviving fragments see the photons of 
the thermal
background always at the same energy.
Then, despite the fact that nuclei injected with energies over the 
photodisintegration threshold lose energy by losing mass, they never 
reach the threshold.
The observed pile-up in the modification factors is due solely to the 
multivalued nature of the energy
at time $t$ as a function of the injection energy: Nuclei injected with 
different energies can arrive with the same energy but with different masses.

It is clear that, except in the region of the pile-up, the modification
factor $\eta$ is less than unity, since $\eta = (E/E_g)^{\gamma-1}$. 
This assertion seems to be in contradiction with the conservation of 
particle number. 
Actually, the conservation of the Lorentz factor implies,
\begin{equation}
\kappa \,E_g^{-\gamma}\,dE_g|_{_\Gamma} = N(E,t)\, dE|_{_\Gamma}
\label{es}
\end{equation}
in accord with the conservation of the number of particles in
the spectrum. Moreover, the condition (\ref{es}) completely determines 
the evolution of the energy spectrum of the surving fragments 
(\ref{espectro}). 
Note that in order to compare the modified and unmodified spectra, with 
regard to  conservation of particle number, one has 
to take into account that the corresponding energies are shifted. As follows 
from (\ref{es}), the
conservation of the number of particles in the spectrum is
given by,
\begin{equation}
\int_{E_{_{\rm th}}}^{E_{\pi_{\rm th}}} N(E,t) dE = 
\int_{E_{g_{\rm th}}}^{E_{g{\pi_{\rm th}}}} \kappa\,E_g^{-\gamma}\ dE_g
\end {equation}
with $E_{_{\rm th}}$ and $E_{\pi_{\rm th}}$ the threshold energies
for photodisintegration, and photopion production processes respectively.

Let us now return to the analysis of Fig. 2
in relation to the depression in the spectrum.
In the case of a nearby iron source, located around  3 Mpc, and 
for injection energies below the multivalued region of the 
function $E (E_g)$, $E$ is clearly less than $E_g$ 
and, as a consequence the depression in the 
modification factor is apparent.  
Then, despite the violence of the photodisintegration process via the 
giant dipole resonance, for nearby sources none of the 
injected nuclei are completely  
disintegrated yielding this unusual depression before the bump.
For a flight distance of 3 Mpc, the composition of the arrival nuclei
changes from $A=50$ (for $\Gamma \approx 10^9$) to $A=13$ (for
$\Gamma \approx 10^{11}$).
However, the most important variation takes place in the region of the bump,
where $A$ runs from 48 to 13, being heavy nuclei of $A=33$ the most abundant.
For propagation distances greater than 10 Mpc one would expect just 
nucleons to arrive for injection energies above $9 \times 10^{20}$ eV. In this 
case the function becomes multivalued below the
photodisintegration threshold and then there is no depression at all.
For an iron source located at 3.5 Mpc, the
depression in the spectrum is almost invisible $({\cal O}(1\%))$, 
in good agreement with the
results previously obtained by Elbert and Sommers using  Monte Carlo
simulation \cite{ES}.

\begin{figure}
%\label{} 
\centering 
\leavevmode\epsfysize=4.8cm \epsfbox{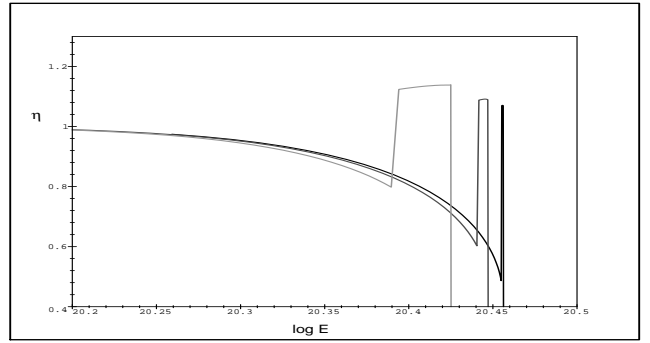}\\ 
\caption{Same as Fig. 4 with spectral index $\gamma = 2.5$.}
\end{figure}

%%%%%%%%%%%%%%%%%%%%%%%%%%%%%%%%%%%%%%%%%%%%%%%%%%%%%%%%%%%%%%%%%%%%%%%%%%%%
\section{Conclusions}
%%%%%%%%%%%%%%%%%%%%%%%%%%%%%%%%%%%%%%%%%%%%%%%%%%%%%%%%%%%%%%%%%%%%%%%%%%%%
We have 
studied the interaction of ultra high energy nuclei with the MBR.
We have
presented a parametrization of the fractional
energy loss for Lorentz factors up to $10^{11}$
that allows us to analyse the evolution of the energy spectrum for 
different nuclei sources.
When considering
an iron source located around
3 Mpc, the spectrum exhibits a depression before a bump not previously 
reported. 
In the light of this finding it is 
tempting to speculate whether the apparent gap in the existing data 
is due to the relative weight of the depression 
and the bump if a source of iron nuclei is responsible 
for the end of the cosmic ray spectrum. This speculation, if true, 
reclaims "botton up" models as a 
possible scenario for the origin of
the highest energy cosmic rays.
The limited statistics in the observed data make
it impossible to resolve the question definitively at this time, and
we are obliged to present this idea as a hypothesis to be tested by experiment.

The  existence of a cutoff or a gap  
which might be present in the observed spectrum
is of fundamental
interest in cosmic ray physics, allowing stringent tests of 
existing models.
The future Pierre Auger Project \cite{Desrep} should provide
enough statistics for a final veredict on these open questions, and in 
particular on the ideas discussed in this paper.

\acknowledgments

Special thanks go to Prof. James Cronin for stimulating discussions.
This research was supported in part by CONICET and FONCYT. L.A.A. thanks FOMEC
for financial support.

\end{document}